\documentclass[useAMS,usenatbib]{mn2e}
\usepackage{epsfig}

\title [Metallicities and ages]
{Metallicities and ages of stellar populations at a 
high Galactic latitude field}

\author[Siegel et al.]
{Michael~H.~Siegel, $^1 \thanks{E-mail:siegel@astro.psu.edu}$
Y\"uksel~Karata\c{s}$^2 \thanks{E-mail:karatas@istanbul.edu.tr}$ and I.~Neill Reid$^3\thanks{E-mail:inr@stsci.edu}$\\
$^1$Pennsylvania State University, 525 Davey Laboratory, State College, PA, 16801, USA\\
$^2$Istanbul University Science Faculty, Department of Astronomy and Space Sciences, 34119 
University-Istanbul, Turkey\\ 
$^3$Space Telescope Science Institute, 3700 San Martin Drive, Baltimore, MD, 21218, USA\\}

\begin{document}

\date{Accepted. Received; in original form}

\pagerange{\pageref{firstpage}--\pageref{lastpage}} \pubyear{2008}

\maketitle

\label{firstpage}

\begin{abstract}
We present an analysis of $UBVRI$ data from the Selected Area SA 141. 
By applying recalibrated methods of measuring ultraviolet excess (UVX),
we approximate abundances and absolute magnitudes for 368 stars over 1.3 
square degrees out to distances over 10 kpc. With the density distribution
constrained from our previous photometric parallax investigations and with
sufficient accounting for the metallicity bias in the UVX method, we are able
to compare the vertical abundance distribution to those measured in previous
studies.
We find that the abundance distribution has an underlying uniform component
consistent with previous spectroscopic results that posit a monometallic thick 
disk and halo with abundances of $[Fe/H]$ = $-$0.8 and $-$1.4, respectively.  
However, there are a number of outlying data points that may indicate contamination
by more metal-rich halo streams.
The absence of vertical abundance gradients in
the Galactic stellar populations and the possible presence of interloping halo 
streams would be consistent with expectations from merger models of Galaxy 
formation. We find that our UVX method has limited sensitivity in 
exploring the metallicity distribution of the distant Galactic halo, owing to the poor
constraint on the $UBV$ properties of very metal-poor stars.  
The derivation of metallicities from broadband $UBV$ photometry remains 
fundamentally sound for the exploration of the halo but is in need of both improved
calibration and superior data.
\end{abstract}


\section{Introduction}

The three-dimensional abundance distribution of the Milky Way field
stars is a key constraint on models of Galactic formation (see, e.g.,
the reviews in Majewski et al. 1993; Freeman \& Bland-Hawthorn 2002). 
The observed abundance distribution is the convolution of the enrichment
history of the Milky Way and the dynamical evolution of its stellar populations
(as reflected in their spatial and kinematical distributions). Specific
models of galaxy formation make specific predictions about the abundance 
distribution measured {\it in situ} from the field stars that can be tested 
against large catalogues of abundance measures.
For example, the hypothesis that the Galactic halo was formed in a global 
collapse was primarily supported by stars on more radial orbits (and therefore
reaching higher $z$ heights above the Galactic midplane) being more metal-poor
than stars on planar orbits (Eggen, Lynden-Bell \& Sandage 1962). However, 
the global collapse theory was unable to account for the lack of an abundance 
gradient in the Milky Way globular clusters (Searle \& Zinn 1978), leading to the
hypothesis that the Galaxy formed at least partly through the merging of larger 
systems -- a hypothesis now well-supported by studies of stellar kinematics and 
spatial distributions (Majewski, Munn \& Hawley 1996; Ivezic et al. 2000; Yanny et al. 2000;
Vivas et al. 2001; Siegel et al. 2002, hereafter S02; Newberg et al. 2002; 
Yanny et al. 2003; Majewski et al. 2003; Rocha-Pinto et al. 2004; Belokurov et al. 2006;
Grillmair et al. 2006; Vivas \& Zinn 2006; Juric et al. 2008)
as well as the properties of globular clusters (see, e.g., Marin-Franch et al. 2008).

The abundance distribution is of particular importance to understanding the
nature of the Galactic thick disk. An abundance gradient in the thick disk 
would favor a scenario in which the thick disk was formed either in the slow late
stages of the early Galactic collapse or the gradual kinematical diffusion of 
disk stars. On the other hand, a monometallic or irregular metallicity distribution
among these stars would favor the thick disk having formed via the kinematical heating of an early thin disk
or directly from merger debris.

The Galactic metallicity distribution is best probed directly through 
spectroscopic surveys (Yoss, Neese \& Hartkopf \ 1987; Allende-Prieto et al. 2006,
hereafter AP06).
However, an alternative method to spectroscopy is the use of field star
$UBV$ photometry (Majewski 1992; Gilmore \& Wyse 1985; Karaali et al. \ 2003). 
$UBV$ photometry allows the measurement of approximate abundances because the
ultraviolet flux of a star is dramatically affected by
the line-blanketing of the star's heavy metals -- a phenomenon first described by Roman \ (1954) and 
Wildey et al. \ (1962). The measurement of ultraviolet excess -- the amount of ``extra" ultraviolet
light produced by metal-poor stars in comparison to metal-rich stars of similar spectral type -- 
is necessarily less precise than spectroscopy. However, it has the advantage of allowing
the simultaneous measurement of many more stars out to the magnitude limit of 
a photometric survey.

The photometric and spectroscopic studies referenced above have
consistently shown vertical abundance gradients 
ranging from $-$0.75 to $-$0.08 dex kpc$^{-1}$, with the gradient
slowly flattening at the high-$z$ regions dominated by the halo. However,
these gradients do not necessarily reflect abundance gradients {\it within} 
any Galactic population. They could also be produced by the transitions
from one mono-metallic population to the other.  Majewski \ (1992), for example, 
argues from a photometric and astrometric survey of the North Galactic Pole 
that the thick disk has no vertical metallicity gradient and a non-Gaussian metallicity 
distribution.  AP06, using a spectroscopic survey of kinematically-selected
stars, demonstrate that neither the thick disk nor the halo have an intrinsic vertical abundance gradient.
They also find that while the halo has a broad range of abundances, the
thick disk has a compact metallicity distribution with a peak metallicity
of $[Fe/H]\sim-0.7$.  Du et al. (2004), using an analysis of BATC photometry, 
posit that {\it all three} old stellar 
populations (old disk, thick disk and halo) lack a vertical abundance gradient.

As this paper was in preparation, Ivezic et al. (2008, hereafter I08) published
an extensive study of $ugr$ photometry taken from the Sloan Digital Sky Survey.
The tremendous number of stars available allows them to probe both broad and 
fine structure in the abundance distribution.  While they confirm
the general properties of the Galactic populations depicted in earlier surveys 
such as AP06, they do so with much more precision.  Remarkably, their results 
would be consistent with the complete absence of a thick disk in the Milky Way 
or its reduction to the Metal-Weak Thick Disk described by Norris (1987).  
Karatas et al. (2008) have also recently analyzed $ugriz$ photometry from 
the CFHTLS DR4.  In contrast to other studies, they find a vertical abundance gradient
in the halo.

In this paper, we present an analysis of stars in the Kapteyn Selected Area SA 141,
which is located near the South Galactic Pole and is ideal for probing vertical 
spatial, kinematical and abundance distributions. Trefzger, Pel \& Gabi (1995)
previously studied the photometric properties of this field in an attempt
to constraint the metallicity distribution. Using 112 stars over 1.9 square 
degrees to a depth of $V=14.8$, Trefzger et al. found evidence for a distinct thick 
disk population and gradients roughly consistent with previous spectroscopic studies.

We now expand upon their result, presenting an analysis of photoelectric $UBVRI$
photometry in a similar field but to greater depth.  The present study provides an 
independent check on the remarkable results of I08 as well as an exploration of 
issues affecting the derivation of metallicities from broadband $UBV$ colours.
Section 2 of this paper describes the observations, corrections for 
foreground extinction and removal of QSOs contamination. Section 3 describes
the method used to measure photometric abundances, absolute magnitudes
and distances while Section 4 shows our analysis of the abundance
distribution perpendicular to the Galactic disk and the comparison of our measures
to those expected from the most recent models.

\section{Observation and data reduction}

\subsection{Photometric Data}

The SA 141 field was observed as part of a larger starcounts survey, described
in detail in S02. In brief, SA 141 was one of twelve fields observed
in the $UBVRI$ passbands with the wide-field Cassegrain CCD camera on
the 1-m Swope telescope at Las Campanas Observatory between 1993 and 1998.
Each field was observed in a uniform grid of pointings. In the case of SA 141, 
the $U$ data cover approximately 1.3 square degrees to a mean depth of $U\sim21$.
Transformation equations to the standard system of Landolt \ (1992)
were derived through matrix inversion techniques and each field
was iteratively transformed to this system. We applied the morphological object
classification techniques of S02 and identified 1294 star-like objects with
measurable $U$ magnitudes.

Since the publication of S02, we have slightly improved the data, generating
astrometry from the IRAF TFINDER program and USNO SA2.0 catalog (Monet al. 1996).
We also combined stars with multiple observations and removed slight photometric 
inconsistencies between the individual pointings using techniques described in 
Siegel et al. \ (2009).

Photometry was de-reddened using the reddening maps and extinction coefficients
of Schlegel, Finkbeiner \& Davis\ (1998).  Although several studies have indicated that the
Schlegel maps over-estimated the reddening (see, e.g., Cambresy et al. 2005), the
stars in the SA 141 field have minimal extinction ($0.011\leq E(B-V)\leq0.024$), minimizing
any impact on our study.

\begin{figure*}
\includegraphics[width=14 cm,clip=]{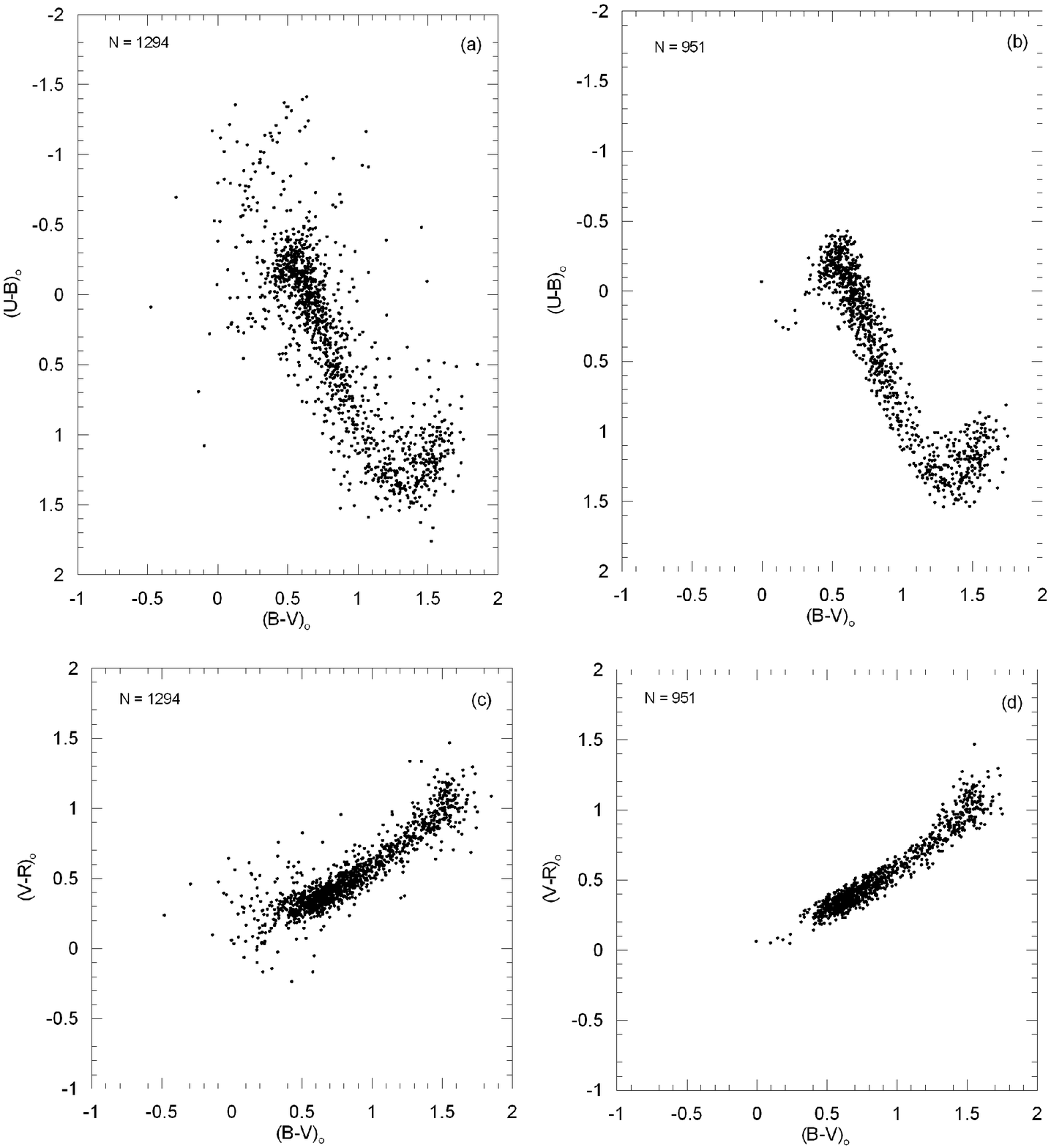}
\caption{Colour-colour diagrams for SA 141 field stars.
Left panels (a) and (c) show $(U-B)_{o}$,$(B-V)_{o}$ and $(V-R)_{o}$,$(B-V)_{o}$ 
two colour diagrams of 1294 stars and star-like objects. Right panels (b) and (d) show $(U-B)_{o}$,
$(B-V)_{o}$ and $(V-R)_{o}$,$(B-V)_{o}$ diagrams for 951 stars after 
possible galaxies and quasars are eliminated.}
\end{figure*}

\subsection{QSO and galaxy contamination}

Deep, high-latitude fields can suffer significant contamination from background
galaxies. While SO2's morphological techniques effectively remove most diffuse 
galaxies, compact galaxies can remain in the sample (Reid et al. 1996; S02). 
Fortunately, $UBV$ photometry provides an excellent secondary resource with which
to remove compact background galaxies.

Figs.~1a and c show the $(U-B)_{o}$,$(B-V)_{o}$ and $(V-R)_{o}$,$(B-V)_{o}$ 
colour-colour diagrams of SA 141. The sample shows a very clear and narrow stellar
locus but has significant contamination from compact galaxies and QSOs, manifested 
in the broader distribution overlying the narrow stellar locus.  To remove these objects,
we applied the selection criterion of Chen et al. \ (2001), which removes objects 
with $(u-g)_0>0.50$. This transforms to a $U-B$ colour of $-$0.48, using the equations
of Smith et al. (2002). We further applied methods outlined in Fan et al. (1999) and
Karaali et al. (2003), which remove objects based on their location relative to the
dominant stellar colour$-$colour locus.

This cleaning reduced the sample from 1294 objects to 951 stars (Figs.~1b and d).
The 343 objects removed from the sample is a far greater than the level of QSO contamination
expected (see, e.g., Richards et al. 2001) and it is likely that our photometric locus method
is removing outlier stars that have compromised measures or unusual properties.  
However, the objects removed from our analysis lie well outside the boundaries of 
the iso-metallicity lines discussed in Sect.~3.3 and their absence does not affect 
our analysis.

\section{Analysis Techniques}

\subsection{Stellar population types}

The $(V_{o}, (B-V)_{o})$ colour-magnitude diagram for the 951 stars in the SA141 sample is plotted in Figure 2. 
We have indicated the cut-off colour indices for thin and thick disks and halo populations
as described by Chen et al. (2001, figure 6).  These cutoffs occur at $(g-r)_0$ colours of 0.33
and 0.20, respectively, which we convert to $(B-V)_0$ colours of 0.54 and 0.40 using the transformation
of Smith et al. (2002).  Our photometric distribution is roughly consistent with Chen, with
the thick disk dominating the starcounts to about $V_0\sim19$
and the halo dominating at fainter magnitudes.

\begin{figure}
\includegraphics[width=8 cm,clip=]{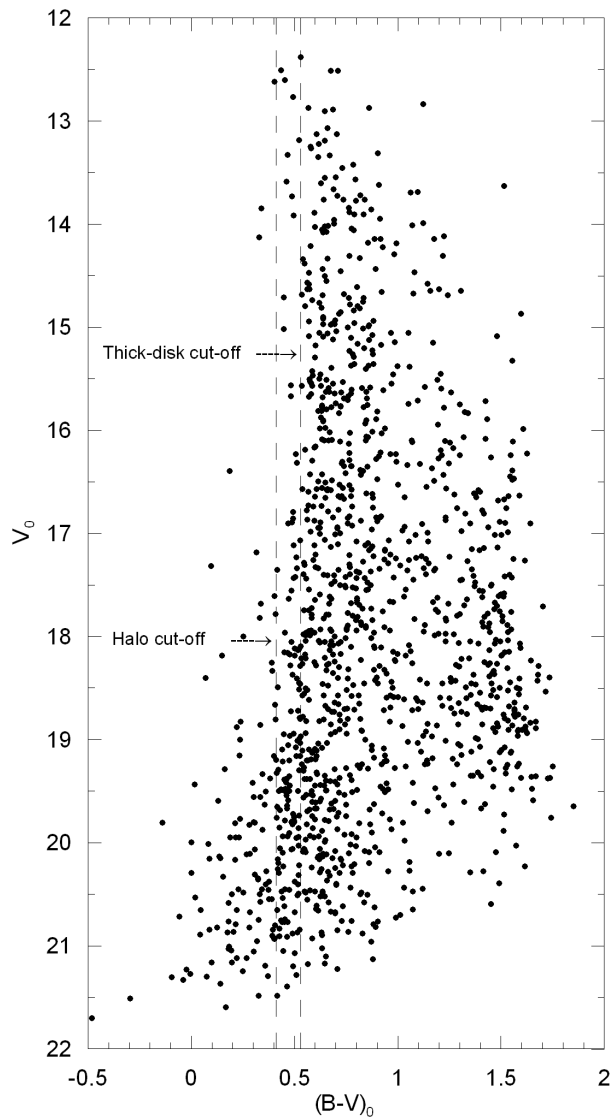}
\caption[] {The $(B$--$V)_o-V_o$ colour-magnitude diagram of 951 stars in SA 141 field.
Dashed lines show the cut-offs of $(B-V)_{o}$ for the thick disk and halo derived 
by Chen et al. (2001).}
\end{figure}

\subsection{Model Comparison}

Untangling stellar populations from raw colour-magnitude diagrams is a difficult
process at best and frequently prone to degenerate solutions, especially when
based upon observations in a single direction.  Sophisticated models, which 
incorporate up-to-date spatial and abundance distributions, luminosity functions
and isochrones, can be used to make Monte Carlo comparisons with which to constrain 
the properties of the underlying stellar populations.  The most readily available 
and recently updated model is the Besancon model of stellar population synthesis 
(Robin et al. 2003, hereafter R03)\footnote{Available online at http://bison.obs-besancon.fr/modele/.},
which can produce synthetic photometric catalogs based on an updated Galactic structure model 
and an input set of magnitude limits and error functions.

Figure 3 compares the Besancon simulation to the SA 141 photometry.  For the simulation, we 
used the updated default structural parameters given by the Besancon simulator.
We simulated errors using exponential distributions fit to the mean error locus of the data (Figure 4).
No {\it a priori} color limits were included and the magnitude limit was set to the mean imaging
limit of the SA141 data ($V=21.2$; $B=20.9$; $U=21.0$).  The imaging limit was defined in S02 as the faintest magnitude
at which stars and galaxies could be distinguished based on DAOPHOT morphological parameters.

\begin{figure*}
\includegraphics[width=14 cm,clip=]{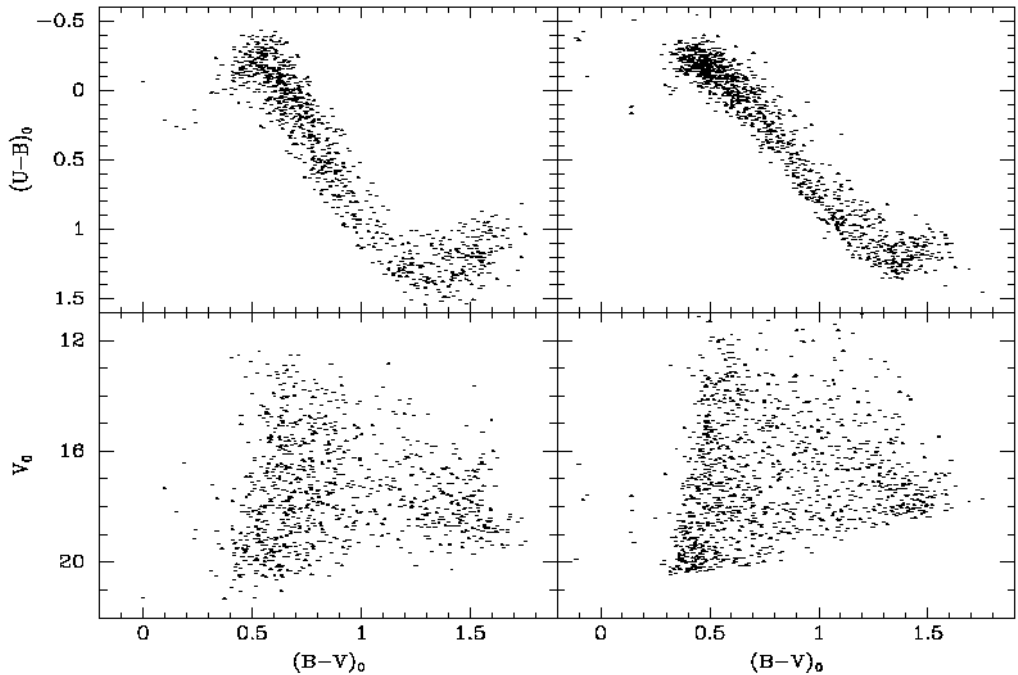}
\caption{A comparison of the SA141 photometry (left panels) to the Besancon simulation (right panels).
 Note the much tighter blue edge population in the simulation.}
\end{figure*}

\begin{figure*}
\includegraphics[width=14 cm,clip=]{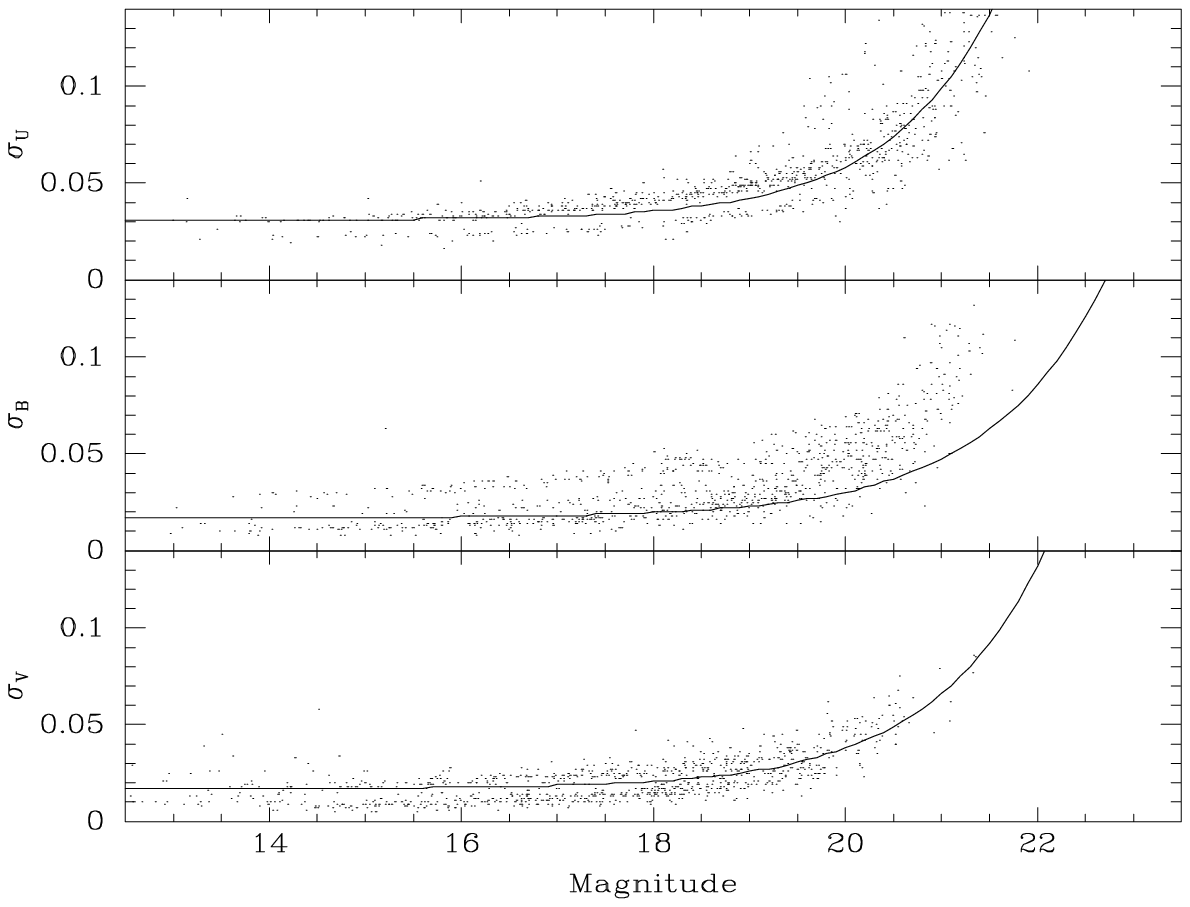}
\caption{Error measurements for stars from the SA141 field.  The error trends show multiple loci
owing to the varying quality of the data.  The solid line represents the error
function of the simulated data.}
\end{figure*}

Both the real and simulated data fall along an arced locus in colour-colour space, 
with the location of each star being a function of its effective temperature and
stellar metallicity (as explored in detail in Sect.~3.3).  The numerous blue stars
that create a clump in upper left corner of the colour-colour diagram would 
correspond to the blue edge depicted in Fig.~2 and shown in the lower panels of Fig.~3.
The blue edge is comprised of intrinsically bright early-type stars.  These stars are faint
in the CMD and therefore several kpc distant in regions dominated by the metal-poor halo.  The
long trail of stars to redward is the increasing contribution of the intrinsically fainter 
(and therefore nearer) late-type stars, which would be in regions of the Galaxy 
dominated by the thick and thin disks.

While the basic features are similar, a number of discrepancies are seen.
The simulated blue edge is much more populous than the observed
blue edge.  The simulated blue edge also has a much tighter photometric
distribution than the real blue edge in both colour-magnitude and colour-colour 
space. Since the Besancon model accounts for photometric error and reproduces 
the redder features of the CMD quite well, it is likely that this discrepancy 
is intrinsic to the model.

The Besancon model uses a halo comprised of a single stellar population which
has a simple power law spatial distribution, a mean abundance of $[Fe/H]\sim-1.78$
and a metallicity dispersion of 0.5 dex. However, the numerous studies referenced 
in Sect.~1 indicate that simple models have difficulty reproducing the halo spatial
distribution and demonstrate that the halo is partly, perhaps mostly, comprised of 
streams of stars stripped from objects during the hierarchical formation of the Milky Way.
The stark difference in uniformity between the real and simulated blue edge argues 
against simple single-population models of the deep halo.

We also note that the simulated galaxy has a slightly different overall colour-colour
distribution, with a gentler slope and a bluer cutoff in the $(U-B)_0$ colour 
distribution. This may not indicate a model deficiency as much as a mismatch between
the real and simulated filters. Our observational data are calibrated to the standards
of Landolt (1992) while the Besancon simulation is tied to the model atmospheres of 
Lejeune, Cuisinier \& Buser \ (1997, 1998), which itself is tied to the observational plane by 
the calibration of Schmidt-Kaler (1982).  It is unsurprising that some discrepancy
would arise between two different calibrations.

\subsection{Metal abundances and absolute magnitudes}

The metal content of main sequence stars can be measured by the amount
of line-blanketing absorption affecting their ultraviolet photometry.
Low metal abundance produces weak atmospheric absorption lines and increased
ultraviolet emission, or ultraviolet excess (UVX; Wildey et al. 1962).

Sandage (1969) detailed a technique 
for using the UVX -- as measured through $UBV$ photometry -- to measure approximate
abundances.
The method determines the two-colour location of a star with reference to 
iso-metallicity ridgelines. The UVX measure, $\delta(U-B$), is the difference
between each star's $(U-B)_0$ colour and that of a metal-rich star of identical
$(B-V)_0$, with the reference population defined in this case by the Hyades 
sequence, which Cayrel, Cayrel de Strobel \& Campbell \ (1985), 
Boesgaard \& Friel \ (1990) and Taylor \ (1998)
estimate has an abundance between $[Fe/H]$=$+$0.10 and $[Fe/H]$=$+$0.13. 
We adopt an abundance of $+$0.13 dex for our analysis.

Figure 5 shows the de-reddened two-colour, $(U$--$B)_{o}$--$(B$--$V)_{o}$ 
of the SA 141 stars that fall within the colour range of the Karatas \& Schuster
(2006) UVX calibration. The solid lines indicate the metal content, ranging for
the reddest line, which is a metal-rich line from the Hyades sequence, to the 
bluest line, which indicates the maximum underabundance envelope defined 
by Sandage (1969).

\begin{figure}
\includegraphics[width=8 cm,clip=]{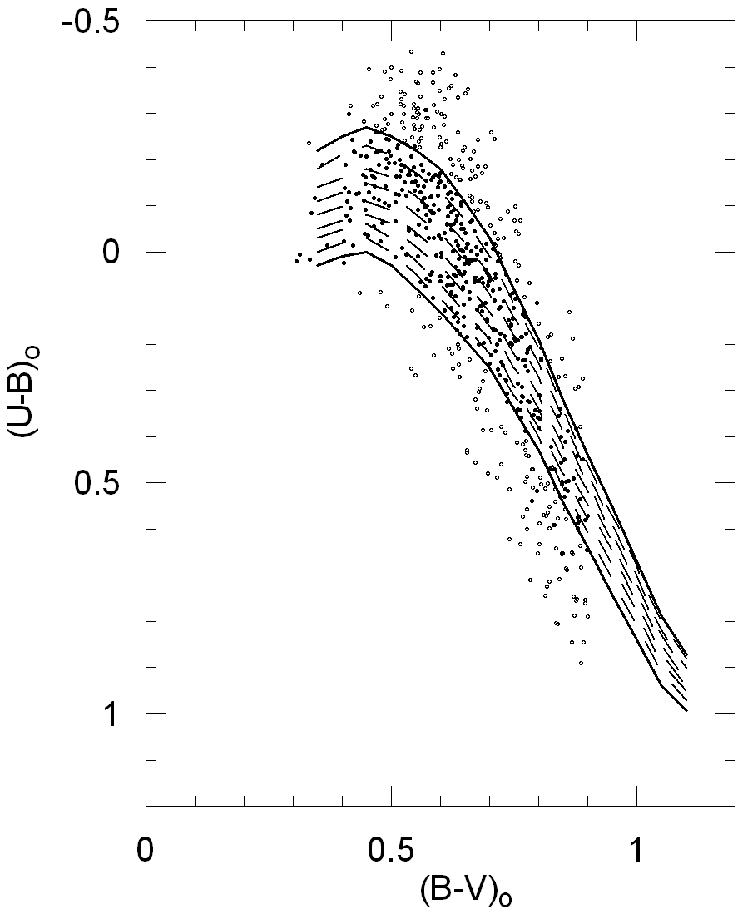}
\caption[] {$(U$--$B)_{o}$-$(B$--$V)_{o}$ two-colour diagram for stars in SA 141 field.
Standard relations for the Hyades and for the metal-poor lines of $(U$--$B)_{0.25M}$, 
$(U$--$B)_{0.50M}$, $(U$--$B)_{0.75M}$, and $(U$--$B)_{M}$ (metal-free) from Sandage (1969) are 
drawn as solid lines. Intermediate metal-poor lines of $(U$--$B)_{0.125M}$, $(U$--$B)_{0.375M}$,
$(U$--$B)_{0.625M}$, $(U$--$B)_{0.875M}$ are drawn as dashed lines.  The decimal fractions
indicate the proportion of the maximum ultraviolet excess, which corresponds to zero
metallicity.  Open points are stars that fall outside of our constrained abundance range.}
\end {figure}

Inspection of Fig.~5 shows that the stars at bluer $(B-V)_0$ colours tend
to lie along the more metal-poor lines while redder stars tend to lie along 
the more metal-rich lines. This is in keeping with expectations of standard
Galactic structure models. The redder stars are intrinsically faint and 
nearby, in regions dominated by the metal-rich thin disk. The bluer stars 
are intrinsically bright and distant, in regions dominated by the metal-poor halo.
A significant number of faint blue halo stars lie beyond the most metal-poor line,
which corresponds to $[Fe/H]$=$-$2.23 in the calibration of Karatas \& Schuster (2006).
This could represent the low-metallicity tail of the halo but would also
represent photometric scatter and evolved stars that have colours outside the 
main sequence locus. A smaller number of nearby red disk stars lie redward of
the Hyades line, possibly representing the high-metallicity tail of the disk.
Excluding the outlying stars from analysis can lead to a systematic bias, which we
address in Sect.~4.2.

To derive $[Fe/H]$ abundances, we first measure $\delta(U-B)$, the colour difference 
between each star and the maximum metallicity line.  This measure is then transformed 
to $\delta_{0.6}$, the UVX measure normalized to a $(B-V)_0$ colour of 0.6 
(where $\delta(U-B)$ reaches its peak value) along the indicated iso-metallicity
ridgelines. Eight constant-abundance lines of Sandage (1969), drawn in Fig.~5, 
have been used to transform the $\delta(U$--$B)$ measures of our stellar sample 
into normalized $\delta_{0.6}$ measures. Constant-metallicity lines are scaled as
a fraction of the maximum line-blanketing "M" value in increments of 0.125.  
The lines are taken from Table 1A of Sandage with adjustments as detailed in 
equations (2)$-$(4) of Karatas \& Schuster (2006).

We convert the $\delta_{0.6}$ measures to approximate abundances using the
Karatas \& Schuster (2006) relation:

\begin{eqnarray}
[Fe/H]=+0.13(\pm0.04)-4.84(\pm0.60)\delta_{0.6}-\nonumber\\
7.93(\pm2.24)\delta_{0.6}^2
\end{eqnarray}

\noindent This relation is valid for the ranges of $+0.41 \leq[Fe/H]\leq -2.23$,
$-0.04\leq \delta_{0.6}\leq +0.29$ and $0.3\leq (B-V)_0 \leq 0.9$.  368 of the SA 141
stars fall within this parameters space.

Absolute magnitudes of program stars are then estimated with the following relation of 
Karatas \& Schuster (2006):

\begin{eqnarray}
M_{V}= 2.77+4.38(B\rm{-}V)_{o}^2-48.98\delta_{0.6}^2 \nonumber\\
~~~~~~~+219.75(B\rm{-}V)_{o}~\delta_{0.6}^2-198.93(B\rm{-}V)_{o}^2~\delta_{0.6}^2\nonumber\\
~~~~~~~+11.04(B\rm{-}V)_{o}^3~\delta_{0.6}
\end{eqnarray}

Abundance uncertainties are calculated by propagating the photometric 
uncertainties through the relevant equations.  The abundance uncertainties
are large, starting at a mean 0.2$-$0.3 dex for nearby ($z<4$ kpc)
stars, reaching 0.5 dex at 8 kpc and 0.7$-$0.8 dex at 15 kpc.  This indicates
that the abundance distribution of the thin and thick disks should be 
well-constrained.  However, the deep halo, where the number of stars
is low and the photometric uncertainties high, may be poorly constrained,
an issue we address in Sect.~4.2.

Our measurement of stellar abundances, absolute magnitudes and distances is
predicated on the assumption that these stars are all main sequence stars.
This is not the case and some degree of contamination from evolved
stars is expected.
The evolved stars will be more distant than the main sequence 
stars and consequently more metal-poor, given the abundance distribution 
identified by previous investigators and constrained in Sect.~4.  
This may cause analysis of the nearby sample to skew metal-poor.

Fortunately, the Besancon simulated galaxy retains information on 
luminosity class. To evaluate the effect of this on our derived abundance
distribution, we ran the simulated galaxy through our analysis pipeline 
and examined UVX-based metallicity measures as a function of midplane height,
both with evolved stars removed from the sample and with evolved stars retained
and evaluated as though they were main sequence stars.  We found that retaining 
the evolved stars caused a small metal-poor skew in the derived metallicity 
that had a maximum of $-$0.1 dex but was generally around a few 0.01 dex.
The reason the effect is so small is that while the evolved stars comprise 
10-20\% of the starcounts at any magnitude, a significant number lie
outside the maximum and minimum UVX boundaries shown in Fig.~5 and are 
rejected on that basis.  Indeed, it is likely that some of the outliers 
in Fig.~5 are, in fact, evolved stars. Additionally, the relatively blue cutoff of our
abundance and absolute magnitude calibrations
removes the redder stars where the luminosity difference between giant
and dwarf is larger and the contamination is therefore from far more distant
and more metal-poor stars.  We have chosen not to remove the evolved star 
bias as it is smaller than the abundance uncertainties.

\section{Metallicity Distribution}

\subsection{Raw Abundance Gradients}

The trend of mean metal abundance $<Fe/H]>$ against $<z>$ height above the
Galactic plane is shown in Figure 6 and detailed in Table 1. Mean metal 
abundances are derived from maximum likelihood Gaussian fits for star 
numbers and $z$ intervals given in Table 1.  As can be seen in Figure 6, at
distance of $z < 4$ kpc -- where the S02 and Robin et al. models indicate that
the thin and thick disks dominate -- there is a vertical abundance gradient of 
approximately $d[Fe/H]/dz = -0.15$ dex $kpc^{-1}$.
At distance beyond 5 kpc -- where the halo begins to dominate -- the metallicity measures are less precise but 
are consistent with a flattening of the abundance gradient (0.03$-$0.06 dex $kpc^{-1}$).
At large distances ($>12$ kpc), the abundance seems to {\it rise} slightly.  
However, the data at these faint magnitudes suffers from low completeness levels 
and it is likely that this reflects a bias near the magnitude limit or contamination
from evolved horizontal branch (HB) stars (Sections~4.2 and 4.3).

\begin{table}
\centering
\caption{Mean metal abundances for $z$ intervals (Column 1).  Column 2 shows the mean $<z>$ of
the stars in each bin while Columns 3 and 4 show the mean $<[Fe/H[>$ and uncertainty of the mean,
respectively. Column 5 indicates the calculated abundance dispersion of the stars while 
Column 6 indicates the number of stars in each bin.}
\begin{tabular}{lrcrrrr}
\hline
$z$ range &$<z>$ &$<[Fe/H]>$ & $\sigma_{<[Fe/H]>}$ &$\sigma_{[Fe/H]}$ &$N$\\
\hline
$[0, 1]$   &  0.64 & -0.50 & $\pm$0.07 & 0.56    &  75  \\
$(1, 2]$   &  1.48 & -0.61 & $\pm$0.06 & 0.55    & 116  \\
$(2, 3]$   &  2.47 & -0.79 & $\pm$0.12 & 0.59    &  43  \\
$(3, 4]$   &  3.53 & -0.93 & $\pm$0.11 & 0.28    &  31  \\
$(4, 5]$   &  4.52 & -0.65 & $\pm$0.17 & 0.48    &  23  \\
$(5, 6]$   &  5.38 & -0.96 & $\pm$0.21 & 0.60    &  18  \\
$(6, 7]$   &  6.52 & -0.95 & $\pm$0.23 & 0.31    &  12  \\
$(7, 8]$   &  7.34 & -0.55 & $\pm$0.24 &  ...    &   9  \\
$(8, 9]$   &  8.52 & -1.23 & $\pm$0.35 &  ...    &   8  \\
$(9, 10]$  &  9.38 & -1.13 & $\pm$0.29 &  ...    &   9  \\
$(10, 11]$ & 10.54 & -0.99 & $\pm$0.33 &  ...    &   7  \\
$(11, 13]$ & 12.31 & -0.27 & $\pm$0.25 &  ...    &   9  \\
$(13, 15]$ & 14.02 & -0.43 & $\pm$0.20 &  ...    &  11  \\
$[15, 21]$ & 16.73 & -0.50 & $\pm$0.39 &  ...    &   7  \\
\hline
\end{tabular}
\end{table}

\begin{figure}
\includegraphics[width=9 cm,clip=]{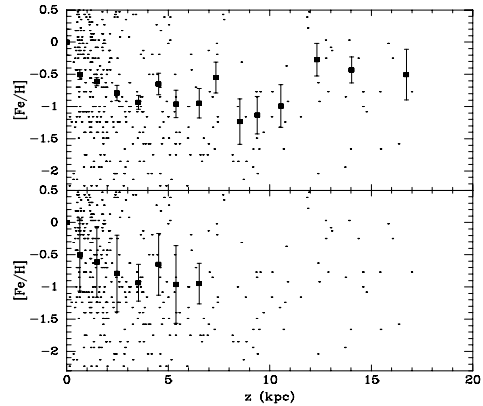}
\caption[]{Relation between $<[Fe/H]>$ and $<z>$ for the range of
$0 \leq z \leq 20$. There is a slight metallicity gradient 
$d[Fe/H]/dz = -0.15$ dex $kpc^{-1}$ for $z < 4$ kpc, where the thin and
thick disks dominate. Error bars in the top panel reflect the formal 
uncertainty in the mean calculated by maximum likelihood from the input photometric errors.  Error bars in 
the bottom panel reflect the measured intrinsic abundance dispersion.}
\end{figure}

\subsection{Debiasing and Deconvolving the Abundance Distribution}

The raw mean abundances shown in Figure 6 can not be taken at face value
due to two complicating factors. First, the vertical abundance distribution 
is a convolution of the intrinsic abundance and  density distributions of 
the component populations. Second, the method itself induces bias.  
Our metallicity calibration only extends from $[Fe/H]=+0.41$ to $[Fe/H]=-2.23$.
This cuts off the low-metallicity tail of the halo and the high-metallicity tail 
of the disk -- a process shown visually in Figure 5. Removing these outlying stars
results in a systematic bias.

The dramatic effect of this bias can be demonstrated by applying our UVX analysis 
pipeline to the Besancon simulated data. Figure 7 compares the simulated abundances with
the UVX-derived abundances.  While the two scales follow each other closely 
at the metal-rich end, the trend veers at low abundances.  The reason is that the 
metal-poor halo stars have a large abundance dispersion -- both intrinsic 
(the Besancon halo has a 0.5 dex dispersion) and observational (0.5$-$0.8 dex from
the $UBV$ photometric uncertainties propagated through the UVX equations).
The low-abundance cutoff removes the (intrinsic plus photometric scatter) 
low-metallicity tail of this distribution, while leaving the high-metallicity tail 
intact. In the faintest reaches of our halo sample, this shifts the derived mean 
abundance by as much as 0.4 dex.

Figure 8 contrasts the underlying intrinsic mean abundance distribution against
the UVX-derived mean abundance distribution.  The difference between the two 
is stark and is the result of the abundance cut-off.  A similar bias occurs for 
the metal-rich thin disk, but is not as dramatic owing to the smaller uncertainties 
for the bright disk stars and the smaller (0.2 dex) intrinsic dispersion of thin
disk abundance.  Note also that the error bars in Figure 8 are smaller than those of Figure 6,
owing to the larger number of faint halo stars in the real data.

\begin{figure}
\includegraphics[width=9 cm,clip=]{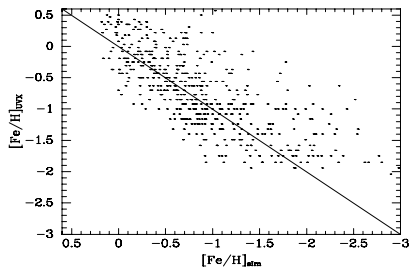}
\caption[]{A demonstration of the bias induced by our UVX analysis.  The points show the input
abundances from the Besancon simulation contrasted against those derived from the 
UVX pipeline and simulated $UBV$ photometry.  The line represents unity.  The sudden shift of the low abundances
away from unity is not the result of miscalculation or error but the increase in 
dispersion combined with the removal of the metal-poor measures whose UVX properties
lie beyond the last ridgeline in Figure 5.}
\end {figure}

\begin{figure}
\includegraphics[width=9 cm,clip=]{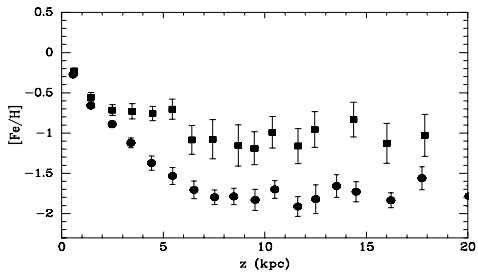}
\caption[]{A demonstration of the metallicity bias induced by UVX analysis.  
The circles represent the intrinsic mean abundance distribution
of the Besancon simulated data.  The square points represent the {\it derived} 
mean abundance measures from the simulated $UBV$ photometry.  In both cases, the uncertainties are calculated
from maximum likelihood (with an assumed uncertainty of $\sigma_{Fe/H}=0.2$ for the intrinsic abundances).  Because the UVX method has large uncertainties at 
faint magnitudes and only extends to $[Fe/H]=-2.23$, a significant metal-rich 
bias is induced when the stars scattered to low abundances are removed, 
as demonstrated in Figure 7.}
\end {figure}

Untangling the halo metallicity distribution under these circumstances is 
beyond the scope of this paper --and may indeed be impossible if the halo 
metallicity distribution is {\it not} a simple single population but, as seems
likely, comprised of multiple components and/or stream of stars.  However, 
since (a) the SA 141 data and UVX calibration are on the same photometric
system (both are tied to the Landolt standards) and; (b) the nature of the 
low-metallicity bias is {\it known}, it is possible to construct toy models 
for comparison.

We constructed synthetic metallicity distributions by assuming the structural
parameters detailed in S02, metallicity distributions from AP06 and R03 and 
a distance-uncertainty relation taken directly from the SA 141 data.  
Both formulations posit a halo and thick disk with no metallicity gradient
and abundance dispersions of 0.3 and 0.5 dex, respectively.  The AP06 mean 
metallicities are $-$0.68 and $-$1.4 dex for thick disk and halo, respectively.
The corresponding R03 abundances are $-$0.78 and $-$1.78 dex.  For the thin disk, 
we took the Besancon metallicity distribution, which is a combination of seven
thin disks in which metallicity is dependent on age.  The mean metallicity of the 
disks varies with age.  However, the overall weighted mean abundance
of the seven disks is $[Fe/H]=-0.08$ and the weighted
weighted 1$\sigma$ is 0.2 dex.  We applied the low- and high-abundance cutoffs 
that are applied to the real data to replicate the bias.

Figure 9 shows the result of these comparisons. What is striking about the 
distributions is that the photometric properties of the Galactic populations
show a rough distribution consistent with the models, but overlayed with several
metal-rich outlying data points (at 4.5, 7.5 and beyond 12 kpc).  S02 
(and many studies cited therein and in Sect.~1) have suggested that the halo 
field star distribution has two components.  The first is a uniform flattened 
metal-poor distribution; the second an irregular mass of star streams created 
during the hierarchical formation of the Milky Way (although Carollo et al. 2007,
using the SDSS, argue for two distinct halo structures, only distinguishable 
beyond 10 kpc).  If the halo did have a uniform structure overlayed with streams, 
the abundance distribution in a pencil-beam survey {\it might} look like Figure 9,
with a uniform distribution occasionally interrupted by more metal-rich populations.
Inspection of Figure 6 shows a hint of this at the outlying points, where the 
individual points seems to show a vague bimodality, although the scatter 
it too large to be certain.

Deep proper-motion and/or radial velocity surveys of SA 141 would be needed 
to test this suggestion.  It remains possible that these data points merely 
reflect some inadequacy of the method.  Nevertheless, if verified, 
it would demonstrate that the stellar streams identified so clearly in I08 
can also be identified in smaller less-precise archival $UBV$ data sets by 
their departure from the underlying uniform distribution.

\begin{figure}
\includegraphics[width=8 cm,clip=]{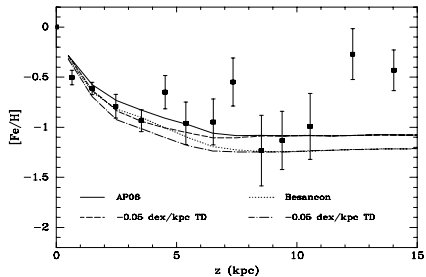}
\caption[]{Comparison of the UVX-based abundances to those predicted by 
the Besancon and AP06 distributions (solid and dotted lines) as well as 
both distributions with a $-$0.05 dex/kpc abundance gradient in the thick disk 
(dashed and dot-dashed lines).  In this figure, the model predictions do not 
correspond to the actual metallicity distribution, but to the {\it measured} 
abundance distribution, taking into account the bias induced by the UVX methodology.}
\end {figure}

If we assume that the metal-rich points are outliers above the general 
distribution, then the literature models successfully reproduce the 
observed abundance distribution.  The Besancon model appears to provide a 
slightly better description of the thick disk while the AP06 model
provides a slightly better description of the halo.  The halo is not
well-constrained enough to address a possible abundance gradient.  
A slight thick disk abundance gradient would bring the AP06
thick disk in line with the UVX results, albeit only because it 
would lower the mean abundance in the narrow range of z heights 
dominated by the thick disk.

\subsection{Young Stars in the Halo?}

Figure 10 shows the $[Fe/H]$, $(B$--$V)_{o}$ diagram overlayed 
with Yonsei-Yale ($Y^{2}$) isochrones of Yi, Kim \&Demarque  \ (2003). 
The turn-off loci from the isochrones are overlayed for ages of 6, 8, 10, 
12, and 14 Gyr, assuming $[\alpha/Fe]$ = +0.30.  While the vast majority of the stars
in SA141 have photometric properties consistent with ages greater than 12 Gyr, a
handful of metal-rich stars are bluer than the old turnoff. These stars are all beyond 
6 kpc and would have implied ages less than 10 Gyr.  A young metal-rich population
of stars is unexpected given previous spectroscopic and photometric
surveys of the halo (see Carollo et al. 2007).

\begin{figure}
\includegraphics[width=8 cm,clip=]{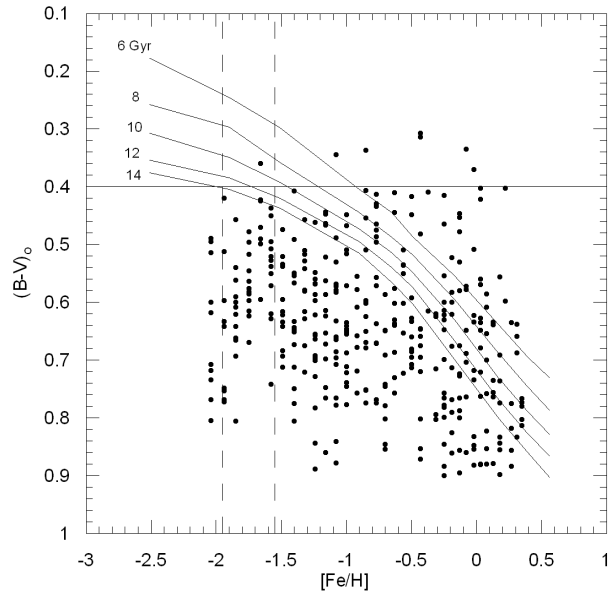}
\caption[]{(a) The $[Fe/H]$, $(B$--$V)_{o}$ plot for the 368 stars in 
SA 141 field. The turn-off loci from the isochrones of Yi et al. \ (2003) 
are plotted for the ages of 6, 8, 10, 12, and 14 Gyr, with $[\alpha/Fe]=+0.30$.
The lines indicate the young metal-poor stars surveyed by Unavane et al.}
\end {figure}

None of the stars comprising this young population have particularly
large photometric errors.  Nor are they concentrated in one part of the SA 141 field, eliminating the possibility 
that they are compromised by bad pixels or nearby saturated stars.  Although
the abundance uncertainties are large for the fainter stars (0.5-0.8 dex) and they
suffer from the aforementioned metal-poor bias, the mean abundance is clearly metal-rich (-0.2 with
a dispersion of 0.7).

These stars are unlikely to represent the halo streams to which we ascribe the outlier points in
Figures 6 and 9.  Halo stream stars are generally thought
to be metal-poor -- as reflected in the blue, metal-poor (BMP) stars described in
Preston, Beers \& Shectman \ (1994), Carney et al. \ (1996) and Unavane, Wyse \& Gilmore (1996).
While some streams, most notably those of the merging Sagittarius dSph, are known 
to be slightly more metal-rich than the halo (Sbordone et al. 2004;
Martinez-Delgado et al. 2005; Chou et al. 2007), our distant SA 141 stars are not
clumped in distance, as would be expected for a tidal stream.

It's possible that these stars are blue straggler stars
(Carney, Latham \& Laird  2005) although there is no reason to believe
that blue stragglers would be so much richer than the canonical halo.
However, the most likely explanation is that these are contaminating HB stars.  
Figure 11 shows the SA~141 $UBV$ diagram with the young stars of Figure 10 
marked as large circles.  To contrast their location against that of HB 
stars, we have generated a synthetic HB from the Dartmouth Stellar 
Evolution Database (Dotter et al. 2007), assuming an abundance of 
$[Fe/H]=-1.5$, [$\alpha$/Fe]=+0.2, an age of 12 Gyr and mass loss 
of 0.05 $\pm$ 0.05 $M_{\circ}$\footnote{The location of the SHB is 
only sensitive to the input parameters at colours bluer than the colour 
range over which we measure UVX abundances.}. It can be seen that the HB 
somewhat overlaps the metal-rich end of MS in colour-colour space.  
If the distant blue metal-poor ``young" stars were on the HB, rather 
than the MS, they would be spread over distances of 25$-$150 kpc, 
a distance range dominated by the extremely metal-poor ($[Fe/H]\sim-2.2$) outer halo
described by Carollo et al. (2007).  We find this the most
likely explanation for these outlying stars.  Spectroscopy
or proper motions of these stars would be able to determine their nature.

\begin{figure}
\includegraphics[width=8 cm,clip=]{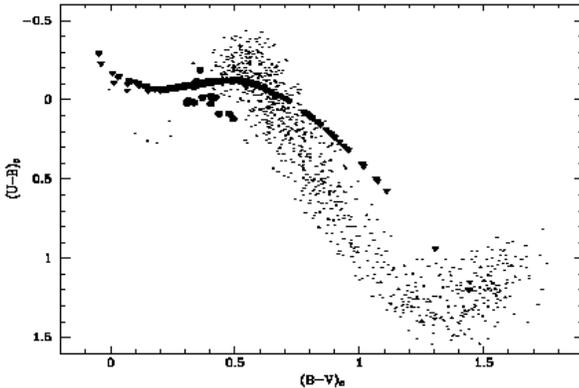}
\caption[]{A analysis of the unusual halo stars.  The dots show the $UBV$ 
photometry of the SA141 field stars. The large circles are the outlying 
apparently young stars from Fig.~9.  The triangles are a synthetic 
horizontal branch.  Note the overlap of the HB and MS sequences where 
the outlying stars are found.}
\end {figure}

\section{Conclusions}

We have performed an analysis of the ultraviolet excesses of stars in 
the SA141 field. We find that photometric uncertainties propagate very 
quickly into large (0.5$-$0.8 dex) abundance uncertainties.  These 
uncertainties can cause systematic biases at the low-metallicity end 
of the UVX calibration.  These biases must be properly accounted for 
when probing the ensemble properties of the field stars.  Specifically,
the limitation of the method to stars more metal rich than $[Fe/H]=-2.23$), when combined with the intrinsic
abundance dispersion of the halo and the necessarily large UVX-abundance 
uncertainties at faint magnitudes, induces a systematic bias in our 
analysis of the population abundance.  This bias can be modeled and 
corrected if assumptions are made about the underlying metallicity 
distribution.

The inferred abundance distribution is roughly consistent with the 
metallicity distribution depicted in I08 as well as the spectroscopic study of AP06
and the simulations of R03. However, there are a number of outlying data 
points that could indicate contamination of the SA 141 field by more 
metal-rich halo streams.

We identify a trace population of apparently
young metal-rich halo stars.  However, these are not associated with
any of the outlier data points and we find it is most likely that
these objects are halo HB stars that have photometric 
properties similar to metal-rich MS stars.

To exploit future $UBV$ data-bases, it will be necessary to refine the 
calibration of the $UBV$ ridgelines based on up-date photoelectric 
data calibrated to the Landolt system as well as to probe the $UBV$
properties of very metal-poor stars to extend the metallicity range of 
the calibration. This would allow the abundance distribution of the distant
halo to be explored with more precision and confidence to test models of 
Galactic formation and unravel the fine structure of the Galactic halo 
and thick disk.

\section{Acknowledgments}
We thank Annie Robin, the referee, for her useful and constructive 
comments concerning the manuscript.

\end{document}